\documentclass{pasj00}
\begin{document}
\SetRunningHead{M. Sasada et~al.}{Multi-wavelength photopolarimetric monitoring
of 3C~454.3 in Dec. 2009}

\Received{2011/04/18}%{yyyy/mm/dd}
\Accepted{2011/12/15}%{yyyy/mm/dd}

\title{Multi-wavelength Photometric and Polarimetric Observations of the Outburst of 3C~454.3 in Dec. 2009}

%%% begin:list of authors
% Do NOT capitalize all letters in "textsc".
\author{
Mahito \textsc{Sasada}\altaffilmark{1},
Makoto \textsc{Uemura}\altaffilmark{2},
Yasushi \textsc{Fukazawa}\altaffilmark{1},
Koji S. \textsc{Kawabata}\altaffilmark{2},
Ryosuke \textsc{Itoh}\altaffilmark{1},
Itsuki \textsc{Sakon}\altaffilmark{3},
Kenta \textsc{Fujisawa}\altaffilmark{4},\altaffilmark{5},
Akiko \textsc{Kadota}\altaffilmark{4},
Takashi \textsc{Ohsugi}\altaffilmark{2},
Michitoshi \textsc{Yoshida}\altaffilmark{2},
Hajimu \textsc{Yasuda}\altaffilmark{1},
Masayuki \textsc{Yamanaka}\altaffilmark{1},
Shuji \textsc{Sato}\altaffilmark{6},
and Masaru \textsc{Kino}\altaffilmark{6}}

\altaffiltext{1}{Department of Physical Science, Hiroshima University, Kagamiyama 1-3-1, Higashi-Hiroshima 739-8526}
\email{sasada@hep01.hepl.hiroshima-u.ac.jp}
\altaffiltext{2}{Astrophysical Science Center, Hiroshima
University, Kagamiyama 1-3-1, Higashi-Hiroshima 739-8526}
\altaffiltext{3}{Department of Astronomy, Graduate School of Science, the University of Tokyo, 7-3-1 Hongo, Bunkyo-ku, Tokyo 113-0033}
\altaffiltext{4}{Graduate school of Science and Engineering, Yamaguchi University, 1677-1 Yoshida, Yamaguchi, Yamaguchi 753-8512}
\altaffiltext{5}{The Research Institute for Time Studies, Yamaguchi University, 1677-1 Yoshida, Yamaguchi, Yamaguchi 753-8511}
\altaffiltext{6}{Department of Physics, Nagoya University, Furo-cho, Chikusa-ku, Nagoya 464-8602}

%%% end:list of authors

%%% Please use the following style in case that sorting by 
%%% affilation is impossible. 
%
% \author{%
%   D-Firstname \textsc{D-Familyname}\altaffilmark{1}
%   E-Firstname \textsc{E-Familyname}\altaffilmark{1,2}
%   and
%   F-Firstname \textsc{F-Familyname}\altaffilmark{2}}
% \altaffiltext{1}{Address of Institute}
% \email{ddddd@xxx.xxx.xx.xx}
% \email{eeeee@xxx.xxx.xx.xx}
% \altaffiltext{2}{Address of Institute
%% `\KeyWords{}' always has to be placed before `\maketitle'.
 %Do NOT move this preamble from here!

\maketitle

\begin{abstract}
 In December 2009, the bright blazar, 3C~454.3 exhibited a strong
 outburst in the optical, X-ray and gamma-ray regions. We performed
 photometric and polarimetric monitoring of this outburst in the optical
 and near-infrared bands with TRISPEC and HOWPol attached to the Kanata
 telescope. We also observed this outburst in the infrared band with
 {\it AKARI}, and the radio band with the 32-m radio telescope of
 Yamaguchi University. The object was in an active state from JD~2455055
 to 2455159. It was 1.3~mag brighter than its quiescent state before
 JD~2455055 in the optical band. After the end of the active state
 in JD~2455159, a prominent outburst was observed in all
 wavelengths. The outburst continued for two months. Our 
 optical and near-infrared polarimetric observations revealed that the
 position angle of the polarization (PA) apparently rotated clockwise by
 240~degrees within 11~d in the active state (JD~2455063---2455074), and
 after this rotation, PA remained almost constant during our
 monitoring. In the outburst state, PA smoothly rotated counterclockwise
 by 350~degrees within 35~d (JD~2455157---2455192). Thus, we detected
 two distinct rotation events of polarization vector in opposite
 directions. We discuss these two events compared with the past rotation
 events observed in 2005, 2007 and 2008.
\end{abstract}

\section{Introduction}
Blazar is a class of active galactic nuclei, whose relativistic jets are
considered to be directed along the line of sight
(e.g. \cite{Blandford79}). Radiation of blazars has three main
properties. First, the blazars emit electromagnetic radiation in a 
broad range from the radio to gamma-ray bands. Their emission consists
of two major components (e.g. \cite{Kubo98}). The low energy component
is synchrotron radiation observed from the radio to the optical,
sometimes extending to the X-ray bands. The high energy component is
inverse-Compton scattering from the X-ray to the gamma-ray
bands. Second, blazars exhibit rapid and violent variability in all
wavelength bands \citep{Antonucci93}. The variability has various
timescales from less than a day (e.g. \cite{Aharonian07};
\cite{Albert07}; \cite{Sasada08}; \cite{Raiteri08a}) to longer than
years (e.g. \cite{Sillanpaa96}). Third, blazars possess relativistic
jets (e.g. \cite{Lister09}), which are responsible for high polarization
observed at optical/near-infrared (NIR), and radio wavelengths. Since the
polarization can be a probe of the magnetic field in the jet,
polarimetric observations are important to study the structure of the
jet.

The temporal variation of the polarization vector is complex in
general. \citet{Jones85} reported that the polarization behavior was
erratic in blazars. On the other hand, several papers reported that the 
polarization vector exhibited systematic variations, for example,
positive correlations between the flux and the degree of polarization
(PD) (e.g. \cite{Smith86}). Recently, \citet{Marscher08} have reported
that the polarization vector in BL~Lac smoothly rotated when the
object was bright. From this result, they proposed that this rotation
indicated an emission zone passing through a helical magnetic field in
the jet. \citet{Abdo10a} reported a rotation of polarization in 3C~279,
and proposed that the rotation event is attributed to a bent
jet. However, there are only a few rotation events which have been
reported to date. 

3C~454.3 is one of the most famous blazars. The object is classified as
Flat Spectrum Radio Quasars (FSRQs), and its redshift is $z=0.859$
\citep{Jackson91}. Although the object had been quiet in the optical
band until 2001, the object has kept showing the active behavior since
then \citep{Villata06}. In 2005, the object showed an exceptional
outburst. In this outburst, the object brightened 
from the radio to the gamma-ray bands (\cite{Fuhrmann06}; \cite{Pian06};
\cite{Giommi06} ; \cite{Villata07}). After this outburst, similar
outbursts were detected in 2007 and 2008. In 2005 and 2007 outbursts,
rotation events of the optical polarization vector were reported by
\citet{Jorstad10} and \citet{Sasada10}. In Dec. 2009, a prominent
outburst was reported for this object in the gamma-ray band by 
{\it Fermi}/LAT and AGILE (\cite{Striani09a}; \cite{Striani09b};
\cite{Escande09}; \cite{Striani10}; \cite{Pacciani10};
\cite{Ackermann10}), in the X-ray band by {\it INTEGRAL}/IBIS
\citep{Vercellone09}, by {\it Swift}/XRT \citep{Sakamoto09}, by 
{\it Swift}/BAT \citep{Krimm09}, in the optical bands
(\cite{Villata09a}; \cite{Bonning09}; \cite{Sasada09}). The object was
the brightest source in the GeV-gamma-ray sky for over a week
\citep{Ackermann10}. The object showed flux variability
over timescales less than three hours and very mild spectral variability
with an indication of gradual hardening preceding major flares. The 
minimum Doppler factor was 13, estimated by using these
results. \citet{Bonnoli11} also estimated the Doppler factors $\sim$~25
during the outburst by constructing a multi-wavelength spectral model.

We performed monitor observations of 3C~454.3 from May 2009 to February
2010 in a multi-color photometric and polarimetric mode using
the Kanata telescope. We also observed in the radio and infrared (IR)
bands. In this paper, we report on the behavior of the 2009 outburst in
these bands, and the detections of two rotation events in the
polarization vector. Directions of these two rotations were different,
suggesting a complex magnetic field in the jet. This paper is arranged
as follows: In section~2, we
present the observation method and analysis in the radio, infrared,
optical and X-ray bands. In section~3, first we report the light curves
and the spectral indexes of the X-ray and optical bands. Then, we report
the temporal behavior of polarization in the optical and NIR
bands. After that, we report the spectral energy distribution from the
radio to optical band. In section~4, we discuss two rotation events
which we detected, by comparing them with the past rotation events
observed in 2005, 2007 and 2008. The conclusion is drawn in section~5.

\section{Observation}
\subsection{TRISPEC and HOWPol attached to the Kanata telescope}
We performed monitor observations of 3C~454.3 using TRISPEC attached to
the Cassegrain focus of the Kanata 1.5-m telescope at Higashi-Hiroshima
Observatory. TRISPEC can perform photometric and polarimetric
observations in the optical and two NIR bands,
simultaneously \citep{Watanabe05}. In the observation of 3C~454.3,
unfortunately, one of the two NIR arrays was not available due to a
readout error. Therefore, we observed the object with the multi-color
photometric and polarimetric monitoring in the $V$ and $J$ bands. A unit
of the observing sequence consisted of successive exposures at four
position angles of a half-wave plate; 0\arcdeg, 45\arcdeg, 22.\arcdeg5,
67.\arcdeg5. A set of polarization parameters was derived from each set
of the four exposures.

We also observed in the multi-color photometric mode in the 
$R_{\rm C}$ and $I_{\rm C}$ bands during the outburst of the object 
using HOWPol (Hiroshima One-shot Wide-field Polarimeter;
\cite{Kawabata08}) attached to a Nasmyth focus of the Kanata
telescope. In this paper, we use multi-band photometric data obtained
with HOWPol in \S~3.3.

The integration time depended on the sky condition and the brightness
of 3C~454.3. Typical integration times were 90 and 85~s in the $V$ and
$J$ bands, respectively. All images were bias-subtracted and
flat-fielded, before aperture photometry. We performed differential
photometry with a comparison star taken in the same frame of
3C~454.3. Its position is R.A.=$\timeform{22h53m58s.11}$,
Dec.=$\timeform{+16D09'07''.0}$ (J2000.0) and its magnitude is
$V$=13.587, $R_{\rm C}$=13.035, $I_{\rm C}$=12.545 and 
$J$=11.858 (\cite{Gonzalez-Perez01}; \cite{Skrutskie06}). After the
differential photometry, we calculated the flux, assuming that the 0~mag
corresponds to the flux with 1.98, 1.42, 0.895 and
0.384$\times$10$^{-5}$ $\rm erg\;cm^{-2}\;s^{-1}$ in the $V$, 
$R_{\rm C}$, $I_{\rm C}$ and $J$ bands (\cite{Fukugita95};
\cite{Bessell98}). 
 
We confirmed that the instrumental polarization was smaller than 0.1~\%
in the $V$ and $J$ bands using the observation of unpolarized standard
stars. We, hence, applied no correction for it. The instrumental
depolarization factors, $\alpha_{\rm Vdep}$ and $\alpha_{\rm Jdep}$, was
derived from the observation using Glan-Taylor prism, to be 
$\alpha_{\rm Vdep}$=0.827 and $\alpha_{\rm Jdep}$=0.928. The observation
was corrected for it. The zero point of the position angle of
polarization (PA) is defined by the standard system (measured from north
to east) by observing the polarized stars, HD~19820 and HD~25443
\citep{Wolff96}.

\subsection{Swift/XRT}
We utilized the archival data of Swift/XRT for deriving the X-ray
light curve. The XRT observations were carried out using the Photon
Counting (PC) readout mode. The XRT data were reduced using FTOOLS in
the HEAsoft package~(v6.6). We extracted the source event within a
radius of 50'' and background event within radii of 80---100''
centered on the source. We used the XSPEC package~(v11.3) to fit the
data. We applied an absorbed power-law model with Galactic absorption
fixed at a value of $N_{H}=1.1\times10^{21}\;{\rm cm}^{-2}$
(wabs*powerlaw model in XSPEC) \citep{Donnarumma09}.

%May 2009 --- Jan. 2010

\subsection{{\it AKARI}}
The NIR spectroscopic observations of 3C~454.3 were carried 
out at 14:42:27 on 12 Dec., 15:25:22 on 13 Dec. and 01:18:31 on 14
Dec. in 2009(UT) with the {\it AKARI} satellite in the framework of the
{\it AKARI} Open Time Observing programs for the Phase 3-II ``Blazar
Variability in near-Infrared, Optical and Gamma-ray regions (BVIOG)''
(PI: M. Sasada). All the observations 
were performed with the spectroscopic mode (AOT IRCZ4; \cite{Onaka10}) 
in which the data were taken with the prism, NP (1.8--5.5~$\mu$m; 
\cite{Ohyama07}), installed in the NIR channel of the Infrared Camera
(IRC; \cite{Onaka07}) of the {\it AKARI} satellite. Each of  
the data reduction procedure, including the subtraction of the detector 
dark current, correction for the high-energy ionizing particles effects,
the shift and co-addition of the exposure frames, and the wavelength 
calibration for NP data, follows those in the IRC Spectroscopy toolkit
for Phase~3 data Version 20110114. In order to correct for the 
sensitivity changes during the phase-3 of {\it AKARI} mission to obtain
the accurate flux level of the spectrum due to the seasonal temperature
fluctuations of 
the detectors, we derived our own system spectral response curve of NP 
by using the spectroscopic datasets of a calibration standard KF06T2 
collected at the nearest epochs of our datasets. In this paper, we used
only 3.0--5.0~$\mu$m data because of a large uncertainty of the flux 
calibration outside this region.

\subsection{Yamaguchi radio telescope}
The radio observation was carried out with the Yamaguchi 32-m radio
telescope at the center frequency of 8.38~GHz and bandwidth of 400~MHz
in the total power mode. The antenna temperatures were measured at the
position of 3C~454.3 and at four positions 2 arc-min (a half of FWHM)
offset to positive and negative in both azimuth and the elevation
directions from the target so as to obtain the true antenna
temperature by correcting the pointing error of the telescope. A flux
calibrator 3C~48 was observed at the same elevation of 3C~454.3. The
flux density of 3C~454.3 was determined from the ratio of the antenna
temperatures of 3C~454.3 to 3C~48, and the flux density of 3C~48 (3.34
Jy, \cite{Ott94}). The accuracy of the measured flux density is
supported to be 5~\% empirically. The observation was ensured to be 5
times from October 21st to December 7th. 

\section{Result}
\subsection{Photometry}
\begin{figure}
  \begin{center}
 \FigureFile(80mm,75mm){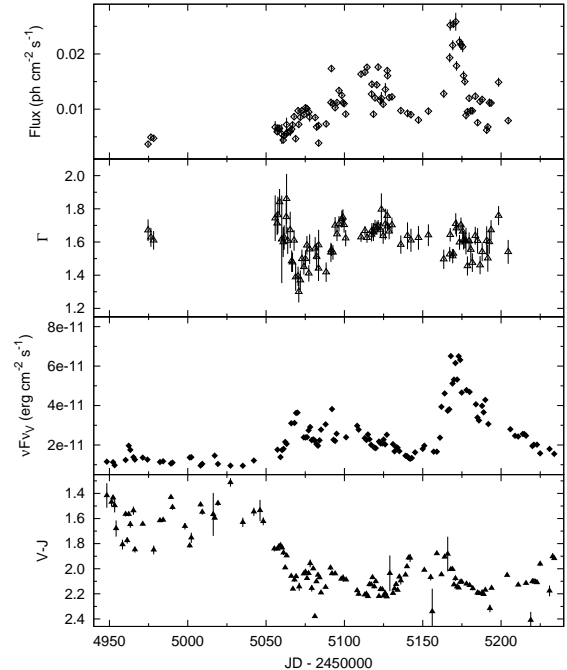}
  \end{center}
  \caption{%
 Light curves, temporal variation of photon index and color variation of
 3C~454.3. From top to bottom, the panels show the light curve and
 temporal variation of photon index $\Gamma$ in the X-ray band at 1~keV,
 the light curve in the $V$ band and $V-J$ color variation.}% 
  \label{fig1}
\end{figure}

Figure~1 shows the light curves in the X-ray and optical bands, temporal
variation of photon index $\Gamma$ and the $V-J$ color variation. 
The X-ray and optical light curves show that the flux of 3C~454.3 was
variable, and we define three states based on the light-curve structure.
The first state is a quiescent state from the start date of our monitoring
to JD~2455055. The second state is an active state from JD~2455055 to
2455159. And the third state is an outburst state after JD~2455159. The
flux in the quiescent state was less variable and faint both in the X-ray
and optical bands compared with the flux after the active state. The
active state was characterized by several short and small flares with
duration of $\sim 10$~d and amplitude of a factor $\lesssim 3$. On
JD~2455159, the object had suddenly become bright both in the X-ray and
optical band, simultaneously. In the decline phase of the outburst, we
can estimate a decline rate, $\tau$, assuming that the flux follows an
exponential decay, that is, $F(t)\propto e^{-t/\tau}$
\citep{Bottcher07}. The decline rates were 14$\pm$2 and 49$\pm$3~d in
the X-ray and optical bands. Hence, the flux decline in the X-ray band
was faster than that in the optical band.

Radio observations at 8.38~GHz were also performed in JD~2455125,
2455127, 2455133 and 2455172 by Yamaguchi radio observatory as mentioned
in \S~2.4. The radio flux density was constant at 8.1~Jy and no
significant change was observed during the active and outburst
states. The lack of a tight correlation between the radio and optical
fluxes was also reported in past studies of 3C~454.3 \citep{Villata07}. 

The photon index in the X-ray band, $\Gamma$ was almost constant
during our monitoring period, except for a possible variations at the
onset of the active state; a temporary softening of spectra can be
seen in $\sim$JD~2455070, while its variation amplitude is
small. \citet{Raiteri11} also analyzed the same XRT data, and reported
that no real changes in $\Gamma$ could be detected from 2008 to
2010. It is also noteworthy that no prominent change in $\Gamma$ was
associated with the outburst state. 

In the quiescent state, $V-J$ was about 1.6. The $V-J$ color in the 
quiescent state was bluer than those in the active and outburst states. 
This feature indicates a, so-called, redder-when-brighter trend. The 
same color behavior was reported in past observations of 3C~454.3
(e.g. \cite{Raiteri08b}; \cite{Sasada10}). It is widely accepted that
this feature appears because an underlying thermal emission in the UV
band, called as Big Blue Bump; BBB, is bluer than the variable
synchrotron emission \citep{Raiteri07}.

\subsection{Polarimetry}
\begin{figure*}
  \begin{center}
 \FigureFile(140mm,75mm){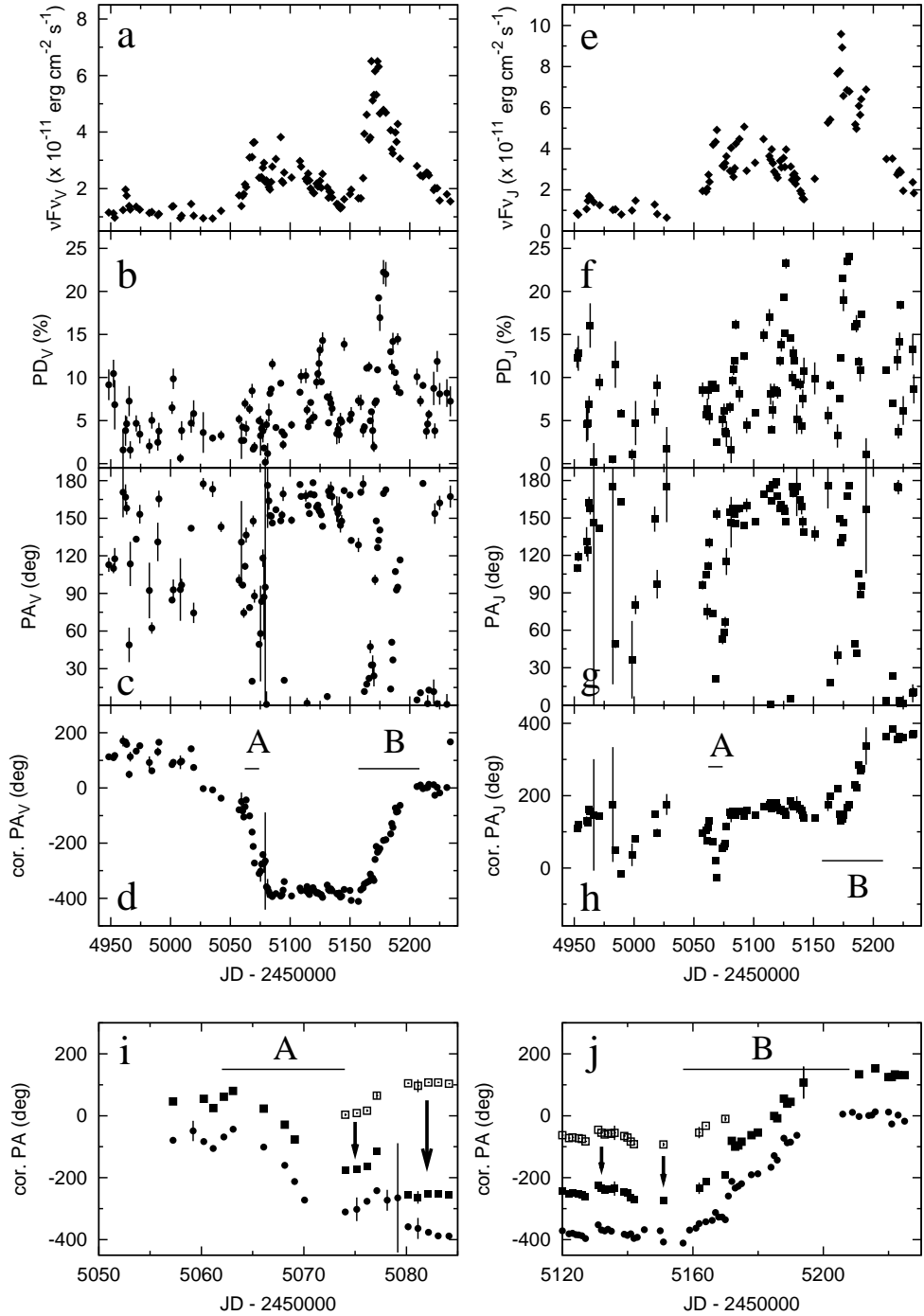}
  \end{center}
  \caption{%
 Temporal variations of the flux and polarization parameters. The left
 and right panels from ``a'' to ``h'' show the each variation in the $V$
 and $J$ bands. The top panels show the light curves of the object. The
 second, third and fourth panels show the temporal variations of the PD
 and PA and the corrected PA. The bottom panels show temporal variations
 of the corrected PA focused on the rotation ``A'' and ``B''. The filled
 circle shows the polarization parameters in the $V$ band, and the open
 and filled squares show the polarization parameters in the $J$ band.}% 
  \label{fig2}
\end{figure*}

Figure~2 shows the light curves and temporal variations of the
polarization parameters in the $V$ and $J$
bands. 
%bands\footnotemark. 
%\footnotetext{These data are available at an
%electrical table. Each column is represented as; Column (1), (2):
%observation date in JD. 
%Column (3), (4): no reddening corrected $V$-band magnitude and
%error. Column (5), (6): $V$-band PD and error. Column (7), (8): $V$-band
%PA and error. Column (9), (10): no reddening corrected $J$-band magnitude and
%error. Column (11), (12): $J$-band PD and error. Column (13), (14): $J$-band
%PA and error.} 
The panel ``b'',
``c'', ``f'' and ``g'' show the temporal variation of the PD and PA in
the $V$ and $J$ bands. The PD in the active and outburst states exhibited
large variations compared with the PD in the quiescent state. The
averaged PD$_{V}$ were 4.5, 6.0 and 8.9~\% and the averaged PD$_{J}$
were 7.3, 9.2 and 12.2~\% in the quiescent, active and outburst states,
respectively. In the outburst state, the maxima of the PD$_{V}$ and
PD$_{J}$ were 22.0$\pm$1.4 and 24.1$\pm$0.4~\% on JD~2455180. The
averaged PD$_{J}$ were higher than the averaged PD$_{V}$ in all
states. The high PD in the $J$ band is partly due to a low
contribution of the unpolarized flux from the BBB component in the $J$
band. 

We corrected the PA assuming that the temporal variation in the PA is
less than 90$^{\circ}$ between neighbor two observations. We defined the
variation as 
$\Delta\;PA_{n}=PA_{n+1}-PA_{n}-\sqrt{{\delta\;PA_{n+1}}^{2}+{\delta\;PA_{n}}^{2}}$, 
where $PA_{n+1}$ and $PA_{n}$ were the n$+$1- and n-th PA and
$\delta\;PA_{n+1}$ and $\delta\;PA_{n}$ were the errors of n$+$1- and
n-th PA. If $\Delta\;PA_{n}<-90^{\circ}(>+90^{\circ})$, we add
$+$180$^{\circ}$ ($-$180$^{\circ}$) to $PA_{n+1}$. If
$|\Delta\;PA_{n}|<90^{\circ}$, we performed no correction of
$PA_{n+1}$. The panel ``d'' and ``h'' show the corrected PA. 

In the panel ``d'', two rotation events can be seen, ``A'' and
``B''. The rotation event ``A'' occurred from JD~2455063 to 2455074 when
the object entered the active state. The rotation event ``B'' occurred
from JD~2455157 to 2455192, during the outburst state. In the quiescent 
state, there was no rotation event. Thus, we can consider that these 
rotation events were associated with the activity of the object. After the
rotation ``A'', the PA was constant during the active state, at about
170$^{\circ}\pm$30$^{\circ}$.

These rotations can be confirmed also in the $J$-band observations,
while the timings of the PA correction are different in several points
as shown in panel ``h''. 
This is mostly due to large errors of PA ($\delta\;PA$) in the
$J$-band observations. The PA correction mentioned above depends on
$\delta\;PA$. In addition, the data number of the $J$-band observation
is smaller than that of the $V$-band one. This is partly due to
mechanical errors of our NIR detector. Thus, the PA correction for the
$V$-band data is more reliable than that for the $J$-band one. For
example, in panels ``i'' and ``j'', we show the corrected PA around
rotations ``A'' and ``B''. The open squares denote the corrected PA of
the $J$-band data. They apparently show different behavior from the
$V$-band PA (the filled circles). However, they becomes consistent if
$-180$ or $-360$~deg is added to the $J$-band PA. It demonstrates that
the $V$- and $J$-band data have the same behavior if the $\pm 180$~deg
ambiguity in PA is taken into account. 

Rotation rates of ``A'' and ``B'' were estimated as $-$26$\pm$2.3 and
$+$9.8$\pm$0.5~deg~d$^{-1}$, calculated by a linear regression model of
PA. If the rotation rate is positive, the rotation direction is
counterclockwise in the $QU$ plane, or the celestial sphere, and vice
versa. Figure~3 shows the temporal variation of the object in the Stokes
$QU$ plane in the $V$ band. The direction of ``A'' was clockwise
and that of ``B'' was counterclockwise in the $QU$ plane, as can be seen
in the top and bottom panels of figure~3. Thus, the directions of these
two rotation were different.

\begin{figure}
  \begin{center}
 \FigureFile(80mm,75mm){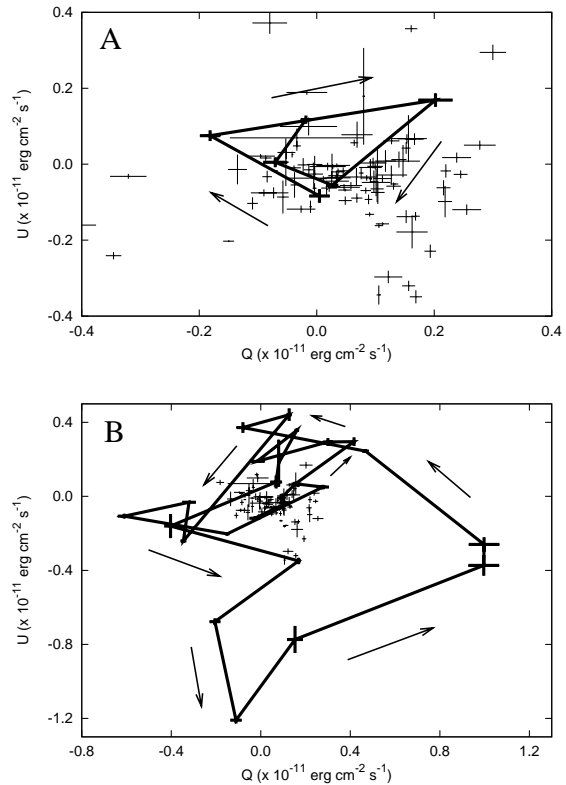}
  \end{center}
  \caption{%
 Behavior of the polarization Stokes parameters on the $QU$ planes
 during the rotation ``A'' and ``B'' in the $V$ band. }%
  \label{fig3}
\end{figure}

Figure~4 shows the light curve and temporal variations of polarization
parameters during the outburst state in the $V$ band. The solid line was
the best-fitted linear function for the corrected PA from JD~2455157 to
2455192. In the fourth panel, we show the residual PA from the linear
function. The PD was low
during the early phase of the outburst. After the outburst maximum, it
became high. The residual PA was largely deviated from zero around
JD~2455170 (the forth panel of figure~4). Around this deviation epoch, 
the $V$ band magnitude was at the maximum (the top panel of figure~4)
and the PD$_{V}$ was minimum ($\sim2$~\%; the second panel of
figure~4).

\begin{figure}
  \begin{center}
 \FigureFile(80mm,75mm){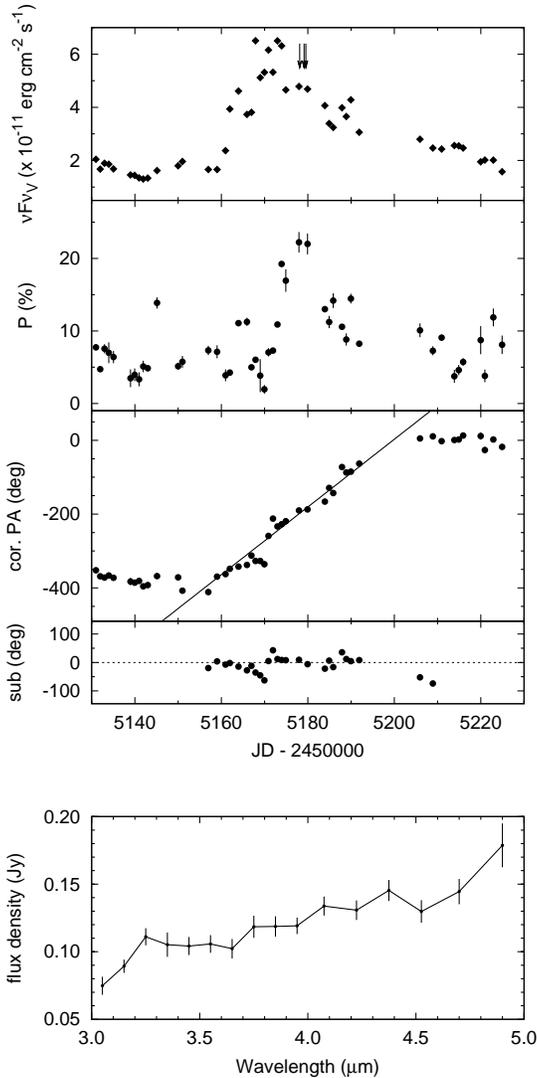}
  \end{center}
  \caption{% 
 Temporal variation of the flux, polarization parameters,
 the PA subtracted from the linear function fitted the data from
 JD~2455157 to 2455192 in the $V$ band and spectrum in the NIR
 region. The arrows in the top panel are the epochs of the {\it AKARI}
 observations. The bottom panel shows the averaged spectrum of three
 spectra.}
  \label{fig4}
\end{figure}

\subsection{Spectrum in the NIR band and spectral energy distribution}
We also obtained the NIR spectroscopic data with {\it AKARI}/IRC during
the outburst state. In the top panel of figure~4, the arrows represent
the observation epochs with {\it AKARI}. The NIR fluxes and
the shape of the spectra were almost identical within the 1-$\sigma$
error level among the three observation epochs. The bottom panel of
figure~4 shows the averaged spectrum of 3C~454.3. The spectrum is
dominated by featureless red continuum emission, indicating that the
synchrotron radiation was dominant in this wavelength range during the
outburst state.

We show the spectral energy distribution (SED) of the object from the 
radio to optical regions during the outburst state in figure~5. The 
NIR data points are the average flux for three epochs obtained
with {\it AKARI}. As for the optical data, the figure includes the
observation with Kanata which was obtained on JD~2455180, closest to the
{\it AKARI} observations. In the radio band, we use public data
observed with the Submillimeter Array (SMA; 1~mm on JD~2455181 and
850~$\mu$m on JD~2455176 ), the the University of Michigan Radio
Astronomy Observatory (UMRAO; 4.8 on JD~2455169.5, 8.0 on JD~2455184.5
and 14.5~GHz on JD~2455179.5) and our data observed with the Yamaguchi
radio telescope (8.38~GHz on JD~2455172). All radio data are obtained
within 10~d of our {\it AKARI} observations. The
SMA data was obtained as part of the SMA flux density monitoring program
and used by permission \citep{Gurwell07}. The optical and NIR data were
corrected for the Galactic interstellar
extinction based on \citet{Schlegel98}. The extinction for the 
{\it AKARI} data were estimated by interpolating the extinctions from 3
to 5~$\mu$m obtained from \citet{Schlegel98}. The solid line represents
the best fitted third-order log-polynomial function. The peak frequency of
the synchrotron component was estimated as
7.6$\times$10$^{12}$~Hz. This is the same order of magnitude as the
values reported in previous studies (e.g. \cite{Raiteri08a};
\cite{Abdo10b}). The optical and NIR data are smoothly connected, which
suggests that at least the emission  from the NIR to optical energy band
can be explained by a synchrotron component.

\begin{figure}
  \begin{center}
 \FigureFile(80mm,75mm){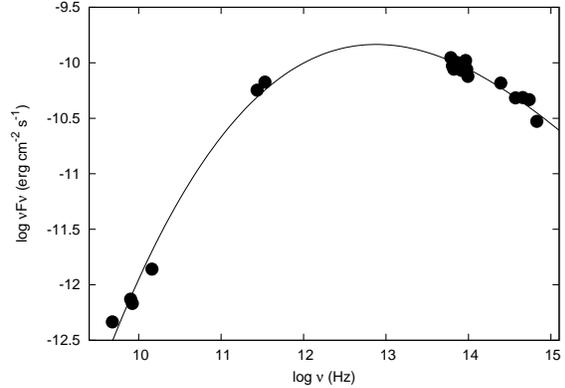}
  \end{center}
  \caption{%
 Spectral energy distribution from the radio to optical bands during the 
 outburst state. The solid line represents the best fitted third-order 
 log-polynomial function.}%
  \label{fig5}
\end{figure}

\section{Discussion}
\subsection{Implication of polarimetric behavior during an outburst state
  in 2009}
Our monitoring observations suggest that two rotation events of polarization
occurred during the active and outburst states of 3C~454.3 in 2009. The
features of the rotation ``A'' and ``B'' were summarized in table~1. The
rotation rates, directions and periods of the rotations are different
between the rotation ``A'' and ``B''.

Two models have recently been proposed for the rotation of polarization;
\citet{Marscher08} and \citet{Abdo10a}. According to \citet{Marscher08},
a rotation of the polarization vector is a sign of helical magnetic
field in the jet. The directions of the rotation events in 3C~454.3 are
both clockwise and counterclockwise. In the case of the simple helical
magnetic field in the jet, the direction of the rotation of polarization
should be one-side. Thus, it needs more complex magnetic field structure
in order to explain the rotation events in 3C~454.3.

\citet{Abdo10a} reported a rotation event in 3C~279, and suggested a
non-axisymmetric structure of the jet, implying a curved trajectory for
the emitting material. This idea can explain rotations in both
directions, and hence, the rotations in 3C~454.3. We can estimate the
distance, $\Delta\;r$, traveled by the emitting material during the
well-sampled rotation ``B'' in 2009. We calculate the travel distance
$\Delta\;r$ as $\Gamma^{2}_{\rm jet}\;c\;\Delta\;t$, where 
$\Gamma_{\rm jet}$ is the bulk Lorentz factor, $c$ is the speed of light
and $\Delta\;t$ is the duration of the rotation event
\citep{Abdo10a}. The duration of the rotation ``B'' was 35~d. In this
case, we adopt a values of $\Gamma_{\rm jet}=19.6$ which was reported by
\citet{Bonnoli11} with the data on JD~2455167.5 (2 Dec. 2009). Thus, we
calculate $\Delta\;r\approx\;3.5\times10^{19}$~cm. The travel distance
of our result was the same order of magnitude compared with that for
3C~279 reported in \citet{Abdo10a}. On the other hand, the duration of
the rotation ``A'' was shorter than that of the rotation ``B''. Thus,
the travel distance of the rotation ``A'' could be shorter than that of
``B'' if $\Gamma_{\rm jet}$ in the active state is same or smaller than
that in the outburst state. 

\citet{Sasada11} reported that there was a positive correlation between 
the amplitudes of the flux and PD of flares in 41 blazars. The
large-amplitude variations in the flux and PD were shown in the outburst
state. Thus, there was a positive correlation between the amplitudes of
the flux and PD in the outburst state of 3C~454.3, which was consistent
with the positive correlation in blazar flares found by \citet{Sasada11}. 

The peak of the PD was delayed by 10~d from the peak of the flux in the
outburst. This behavior might be explained by the following simple
geometrical effect scenario. \citet{Laing80} suggest that the PD is the
highest when the line-of-sight is parallel to the shock plane in the
co-moving frame. In this assumption, the observed flux from the jet is
low because of a low Doppler factor. When the line-of-sight is
perpendicular to the shock plane, the observed flux is high because of a
high Doppler factor, and PD is low because the magnetic field direction
is not aligned. Hence, in this scheme, if the shock comes toward us just
after its formation then the shock plane gradually inclines with respect
to line-of-sight, the PD rises up after the flux peak. It should be
noted that the Doppler factor is changed in this scenario. The Doppler
factor during the 2009 outburst was estimated with multi-wavelength
spectral models, which indicate that the Doppler factor $\delta$ was almost
constant, $\delta\sim25-28.5$ during the outburst (\cite{Ackermann10};
\cite{Bonnoli11}). The observed flux, $F_{\nu,\;{\rm obs}}$, from the
shocked region depends strongly on $\delta$, as
$F_{\nu,\;{\rm obs}}\propto\delta^{3+\alpha}$, where $\alpha$ is a spectral
index. During the period of rotation ``B'' (JD~2455157--2455192), the
maximum and minimum $V$-band flux were 6.501 and 1.658 $\times
10^{-11}\;{\rm erg}\;{\rm cm}^{-2}\;{\rm s}^{-1}$, indicating the flux
changed by a factor of 3.9. However, the amplitude of the flux variation
should be larger than that, because the $J$-band flux variation was
larger than that in the $V$ band. If we explain all the flux variation
only by the variation in $\delta$, $\delta$ should be changed at least
by a factor of 1.4, assuming the spectral index, $\alpha=1$. This is,
however, apparently inconsistent with the result from SED modeling that
it was constant within $\delta\sim 24.5$--$28.5$. Thus, it would be
difficult to explain the observation only by the variation in
$\delta$. More detailed and complex jet modeling would be needed for
interpreting the behaviors of the flux and polarization in the outburst
state.

\subsection{Comparison in five rotation events}
Other rotation events were also reported in active states of 3C~454.3
in 2005 and 2007. \citet{Jorstad10} reported that the optical outbursts
in 2005 autumn and in 2007 were accompanied by systematic rotation of
the PA. \citet{Sasada10} also reported a rotation event in the 2007
outburst.

\begin{figure}
  \begin{center}
 \FigureFile(80mm,75mm){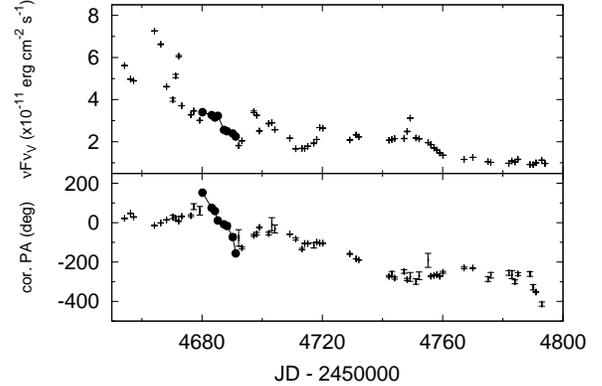}
  \end{center}
  \caption{%
 Temporal variations of the corrected PA in 2008. The top panel shows
 the light curve in the $V$ band. The bottom panel shows the temporal
 variation of the corrected PA in the same band. Filled circles show the
 rotation event in 2008.} 
  \label{fig6}
\end{figure}

In 2008, 3C~454.3 had an outburst in the radio, optical, X-ray and
gamma-ray bands (\cite{Abdo09} ; \cite{Villata09b}). We also monitored
the object during the 2008 outburst with Kanata/TRISPEC in multi-color
photometric and polarimetric mode. Figure~6 shows the light curve and
temporal variation of the corrected PA in the $V$ band in 2008 and
2009. The object reached its outburst maximum on $\sim$JD~2454664. The
object stayed active even after the maximum, until $\sim$JD~2454760, and
then it returned to quiescence. In the lower panel of figure~6, we can
see a gradual decreasing trend of the PA just after the outburst
maximum, which apparently continued until the object returned to 
quiescence. In addition, there is a sign of a rapid and short rotation
of polarization between JD~2454680 and 2454691, as indicated with the
filled circle in figure~6.

\begin{table}
\begin{center}
\caption{%
 Rotation rates in each rotation event.}
\begin{tabular}{llll}
\hline
Event & Rate & Total & rotation period\\
 & (${\rm deg}\;{\rm d}^{-1}$) & (deg) & JD$-$2450000\\
\hline
2005           & $8.7   \pm 1.1$ & --- & ---\\
2007           & $22.0  \pm 3.0$ & 130 & 4333---4338 \\
2008           & $-27.0 \pm 2.0$ & 300 & 4680---4691 \\
2009 ``A''     & $-26.0 \pm 2.3$ & 270 & 5063---5074 \\
2009 ``B''     & $9.8   \pm 0.5$ & 350 & 5157---5192 \\
\hline
\multicolumn{4}{c}{\footnotesize{The 2005 and 2007 rotation events were
 reported by }}\\
\multicolumn{4}{c}{\footnotesize{\citet{Jorstad10} and \citet{Sasada10}.}}\\
\end{tabular}
\end{center}
\label{table1}
\end{table}

To date, five rotation events of polarization were reported in 3C~454.3
during its active and outburst states in 2005, 2007, 2008 and 2009. We
summarize their rotation rates, total degree of the rotation in
table~1. We also show the periods of the rotation events, which we used
for estimating the rotation rates. Three rotation rates are positive
values and two rates are negative. Thus, there are not only
counterclockwise, but also clockwise directions of the PA
rotations. Another feature of the rotation events is that there seems to
be two types of events; fast rotation events (2007, 2008 and 2009 ``A'')
and slow rotation events (2005 and 2009 ``B''). The fast ones have the
absolute rate of the rotation larger than 20~${\rm deg\;d^{-1}}$,
whereas the slow ones have the value smaller than 
10~${\rm deg\;d^{-1}}$. Time durations of two types of rotations are
different. The fast ones have short durations, less than 11~d, and the
slow rotation ``B'' in 2009 has a long duration, 35~d. It could partly
be an observational bias because it is difficult to detect short and
slow rotation events. However, it is worth nothing that there was not
any long and fast rotation from 2007 to 2009.

One possibility is that the fast rotation is the result of an erratic
behavior of the polarization vector. \citet{Villforth10} reported that
the polarization vector can be separated into two components: an optical
polarization core and chaotic jet emission. If the chaotic polarization
component with a short timescale moves around the origin of the $QU$
plane, a short fast rotation can occur. \citet{Ikejiri11} also discussed
that a rotation episode indicated only by a few data points cannot be
distinguished from results of a random walk in the $QU$ plane.

\section{Conclusion}
We performed the photometric and polarimetric monitoring of 3C~454.3 during
2009---2010. The object showed the active and outburst states in the
optical and near-infrared light curves and two rotation events in the
temporal variations of polarization. In the past and our polarimetric
monitoring, five rotation events were reported in 3C~454.3. The
rotations were observed in both clockwise and counterclockwise
directions in the $QU$ plane. The complex model of the structure of the
magnetic field in the jet is needed to explain these rotation events. 
\\
\\
This work was partly supported by a Grand-in-Aid from the Ministry of
Education, Culture, Sports, Science, and Technology of Japan
(22540252,). This research has made use of data from the University of 
Michigan Radio Astronomy Observatory which has been supported by the 
University of Michigan and by a series of grants from the National Science 
Foundation, most recently AST-0607523. The Submillimeter Array is a
joint project between the Smithsonian Astrophysical Observatory and the
Academia Sinica Institute of Astronomy and Astrophysics and is funded by
the Smithsonian Institution and the Academia Sinica. M.~Sasada and
M.~Yamanaka have been supported by the JSPS Research Fellowship for
Young Scientists.

\end{document}